\documentclass[11pt]{article}

\usepackage{amsfonts}
\usepackage{amssymb}
\usepackage{amsmath}

\newcommand{\ket}[1]{|#1\rangle}

\newcommand{\bra}[1]{\langle#1|}

\newcommand{\bracket}[2]{\langle#1|#2\rangle}

\def\half{\textstyle{\frac{1}{2}}}
\def\01{\{0,1\}}

\def\e{\varepsilon}
\def\d{\delta}
\def\a{\alpha}
\def\b{\beta}

\def\ket#1{\mbox{$| #1 \rangle$}}

\def\01{\{0,1\}}

\def\e{\varepsilon}
\def\d{\delta}
\def\a{\alpha}
\def\b{\beta}

\def\ni{\noindent}

\def\ee{\vspace*{3mm}}

\def\loud#1{\noindent{\bf #1 }}
\def\IP{\mbox{\it IP\/}}
\def\EQ{\mbox{\it EQ\/}}
\def\UIP{U_{\mbox{\scriptsize\it \!IP}}}
\def\UIPd{U_{\mbox{\scriptsize\it \!IP}}^{\dag}}
\def\UEQ{U_{\mbox{\scriptsize\it \!EQ}}}

\newcommand{\qed}{\hfill\rule{2mm}{3mm}}

\newtheorem{theorem}{Theorem}
\newtheorem{lemma}[theorem]{Lemma}

\newtheorem{definition}{Definition}

\begin{document}

\title{\Large\bf
A quantum Goldreich-Levin theorem \\ with cryptographic applications}

\author{Mark Adcock\thanks{Email: {\tt mark.adcock@cdcgy.com}}
\hspace{0.3in}
Richard Cleve\thanks{Email: {\tt cleve@cpsc.ucalgary.ca}.
Partially supported by Canada's NSERC.}\\[3mm]
Department of Computer Science\\
University of Calgary\\
Calgary, Alberta, Canada T2N 1N4
}

\date{}

\maketitle

\begin{abstract}
\ni We investigate the Goldreich-Levin Theorem in the context 
of quantum information.
This result is a reduction from the computational problem of 
inverting a one-way function to the problem of predicting a 
particular bit associated with that function.
We show that the quantum version of the reduction---between 
quantum one-way functions and quantum hard-predicates---is 
quantitatively more efficient than the known classical version.
Roughly speaking, if the one-way function acts on $n$-bit 
strings then the overhead in the reduction is by a factor 
of $O(n/\e^2)$ in the classical case but only by a factor 
of $O(1/\e)$ in the quantum case, where $\half + \e$ is the 
probability of predicting the hard-predicate.
Moreover, we prove via a lower bound that, in a black-box 
framework, the classical version of the reduction cannot 
have overhead less than $\Omega(n/\e^2)$.

We also show that, using this reduction, a quantum bit commitment 
scheme that is perfectly binding and computationally concealing 
can be obtained from any quantum one-way permutation.
This complements a recent result by Dumais, Mayers and Salvail, 
where the bit commitment scheme is perfectly concealing and 
computationally binding.
We also show how to perform {\em qubit\/} commitment by a similar 
approach.
\end{abstract}

\section{Introduction}\label{intro}

Fast quantum algorithms are potentially useful in that, if quantum 
computers that can run them are built, they can then be used to 
solve computational problems quickly.
Algorithms can also be the basis of {\em reductions\/} between 
computational problems in instances where the underlying goals 
are different from fast computations.
For example, reductions are often used as indicators that certain 
problems are computationally hard, as in the theory of 
NP-completeness (see \cite{GJ79} and references therein).
Another domain where reductions play an important role is 
in complexity-based cryptography, where a reduction can show that 
breaking a particular cryptosystem is as difficult (or almost as 
difficult) as solving a computational problem that is presumed 
to be hard.

We investigate such a cryptographic setting where quantum 
algorithms yield different reductions than are possible in the 
classical case: the so-called Goldreich-Levin Theorem \cite{GL89}.
This result is a reduction from the computational problem of 
inverting a one-way function to the problem of predicting a 
particular hard-predicate associated with that function.
Roughly speaking, a {\em one-way function\/} is a function 
that can be efficiently computed in the forward direction but 
is hard to compute in the reverse direction, and a 
{\em hard-predicate\/} of a function is a bit that can be 
efficiently computed from the input to the function and yet 
is hard to estimate from the output of the function.
We show that the quantum version of the reduction is quantitatively 
more efficient than the known classical version.
Moreover, we prove via a lower bound that, in a black-box framework, 
the classical version of the reduction cannot be made as efficient 
as the quantum version.

Goldreich and Levin essentially showed that, for a problem instance 
of size $n$ bits, if their hard-predicate can be predicted with 
probability $\half + \e$ with computational cost $T$ then the one-way 
function can be inverted with computational cost $O(T\, D(n,\e))$, 
where $D(n,\e)$ is polynomial in $n/\e$.
Taken in its contrapositive form, this means that, if inverting the 
one-way function requires a computational cost of $\Omega(T)$, then 
predicting the hard-predicate with probability $\half + \e$ requires 
a computational cost of $\Omega(T / D(n,\e))$.
Note that if we start with a specific lower bound of $\Omega(T)$ for 
inverting the function then we end up with a weaker lower 
bound---by a {\em dilution factor\/} of $D(n,\e)$---for breaking 
the hard-predicate.
In \cite{Gol99}, it is shown that the dilution factor can be as 
small as $O(n/\e^2)$.

We show that there is a quantum implementation of the reduction where 
the dilution factor is only $O(1 / \e)$.
We also show that $\Omega(n / \e^2)$ is a lower bound on the dilution 
factor for any classical implementation of the reduction in a 
black-box framework.
In the standard parameterization of interest in cryptography, $T$ is 
assumed to be superpolynomial in $n$ and $\e \in 1/n^{O(1)}$.
In this case, although $1/\e$ is smaller than $n / \e^2$, 
the diluted computational cost, $T / D(n,\e)$, remains superpolynomial 
in both cases.
However, there are other parameterizations where the difference 
between the achievable quantum reduction and best possible classical 
reduction is more pronounced.
One example is the case where $T = n^3$ and $\e = 1/n$.
If we start with a classical one-way function that requires a 
computational cost of $\Omega(n^3)$ to invert and apply the 
Goldreich-Levin Theorem to construct a classical hard-predicate 
then the reduction implies only that the computational cost of 
predicting the predicate with probability $\half + \frac{1}{n}$ is 
lower bounded only by a {\em constant}.
However, if we start with a {\em quantum\/} one-way function that 
requires a computational cost of $\Omega(n^3)$ to invert and apply 
our quantum version of the Goldreich-Levin Theorem then the 
computational cost of predicting the predicate with probability 
$\half + \frac{1}{n}$ is lower bounded by $\Omega(n^2)$.

A particular application of hard-predicates is for bit 
commitment.
Recall the now well-known result that an information 
theoretically secure bit commitment scheme cannot be based 
on the information-theoretic properties of quantum devices 
alone \cite{LC97,May97}.
Of course, this is also the case with {\em classical\/} devices, 
though {\em computationally secure\/} bit commitment schemes have 
been widely proposed, investigated, and applied.
Such schemes can be based on the existence of one-way permutations.
Most of these proposed one-way permutations are hard to invert only 
if problems such as factoring or the discrete logarithm are hard, 
and are insecure against quantum computers, which can efficiently 
solve such problems \cite{Sho94}.
Recently, Dumais, Mayers and Salvail considered the possibility of 
{\em quantum\/} one-way permutations \cite{DMS00}, and showed how to 
base quantum bit commitment on them (see also \cite{CLS01}).
Their scheme is perfectly concealing and {\em computationally binding}, 
in the sense that changing a commitment is computationally hard if 
inverting the permutation is hard.
We exhibit a complementary quantum bit commitment scheme that is 
perfectly binding and computationally concealing.
As with hard-predicates, the dilution factor in the measure of 
computational security is lower than possible with the corresponding 
classical construction.
Furthermore, a possible advantage of our protocol is that the 
information that must be communicated and stored between the 
parties consists of $O(n)$ classical bits for bit commitment 
(and $O(n)$ classical bits plus one qubit for qubit commitment), 
whereas the scheme in \cite{DMS00} employs $O(n)$ qubits.

The organization of this paper is as follows.
In Section~\ref{query}, we investigate a simple black-box problem 
that is related to the Goldreich-Levin Theorem.
In Section~\ref{hard_predicates}, we give definitions pertaining to 
one-way permutations and hard-predicates (classical and quantum 
versions) and
investigate the complexity of reductions from the former to the latter 
(applying results from Section~\ref{query}).
In Section~\ref{bit_commitment}, we show how to use the Goldreich-Levin 
Theorem to construct a perfectly binding and computationally concealing 
quantum bit commitment scheme from a quantum one-way permutation.

\section{A black-box problem}\label{query}

Our results about the Goldreich-Levin Theorem (which are in 
Section~\ref{hard_predicates}) are based on the query complexity of 
the following black-box problem, which we refer to as the {\em GL problem} 
(see, e.g., \cite{Bellare}).
Let $n$ be a positive integer and $\e > 0$.
Let $a \in \01^n$ and let information about $a$ be available only 
from inner product and equivalence queries, which are defined below 
in the classical case (and later on generalized to the case of quantum 
information).

\begin{definition}\rm\label{classical_IP}
A {\em classical inner product (IP)\/} query (with bias $\e$) 
has input $x \in \01^n$ and outputs a bit that 
is slightly correlated with $a \cdot x$ (the inner product 
of $a$ and $x$ modulo two) in the sense that 
\begin{equation}\label{advantage}
\Pr_x[\IP(x) = a \cdot x] \ge \half + \e.
\end{equation}
The above probability is with respect to a random%
\footnote{Unless otherwise specified, a ``random'' element of a 
set means with respect to the uniform distribution.}
$x \in \01^n$.
\end{definition}
\begin{definition}\rm\label{classical_EQ}
An {\em quantum equivalence (EQ)} query has input $x \in \01^n$, 
and the output is 1 if $x=a$ and 0~otherwise.
\end{definition}
The goal is to determine $a$ with a minimum number of $\IP$ 
and $\EQ$ queries.
A secondary resource under consideration is the number of 
auxiliary bit/qubit operations.
It should be noted that, when $\e = \half$, this is essentially 
equivalent to a problem that Bernstein and Vazirani \cite{BV93} 
considered, where $\IP$ queries return $a \cdot x$ on input $x$.
For this problem, $n$ $\IP$ queries are necessary and sufficient 
to solve it classically; however, it can be solved with a single 
(appropriately defined) quantum $\IP$ query.
(See also \cite{TS97}.)
When $\e$ is small---say, $\e \in 1/n^{O(1)}$---an efficient 
classical solution to this problem is nontrivial.
The correctness probability of an $\IP$ query for a particular 
$x$ cannot readily be amplified by simple techniques such as 
repeating queries; for some $x$, $\IP(x)$ may always be wrong.
Goldreich and Levin \cite{GL89} were the first to (implicitly) 
solve this problem with a number of queries and auxiliary 
operations that is polynomial in $n/\e$---and this is the basis 
of their cryptographic reduction in Theorem~\ref{classical_gl}.

We show that any classical algorithm 
solving the GL problem with constant probability must make 
$\Omega(n/\e^2)$ queries (for a reasonable range of values of $\e$), 
whereas there is a quantum algorithm that solves the GL problem 
with $O(1/\e)$ queries.
For the quantum version of the GL problem, quantum $\IP$ and 
$\EQ$ queries are defined 
(in Definitions~\ref{quantum_IP}~and~\ref{quantum_EQ}) 
as unitary operations that correspond to 
Definitions~\ref{classical_IP}~and~\ref{classical_EQ} in a 
natural way.
We begin with the classical lower bound.

\newpage

\begin{theorem}\label{classical_lb}\sl
Any classical probabilistic algorithm solving the GL problem 
with success probability $\d > 0$ requires either more than 
$2^{n/2}$ $\EQ$ queries or $\Omega(\d n / \e^2)$ $\IP$ queries 
when $\e \ge \sqrt{n}2^{-n/3}$.
\end{theorem}

\loud{Proof:}
The proof uses classical information theory, bounding the 
conditional mutual information about an unknown string that 
is revealed by each $\IP$ query, in conjunction with an 
analysis of the effect of $\EQ$ queries.

It is useful to consider an algorithm to be {\em successful\/} on 
a particular input if and only if it performs an $\EQ$ query 
whose output is 1 (at which point the value of $a$ has been 
determined).

We begin by showing that it is sufficient to consider algorithms 
(formally, decision trees) that are in a convenient simple form.
First, by a basic game-theoretic argument \cite{Yao83}, it 
suffices to consider deterministic algorithms, where their input 
data---embodied in the black-boxes for $\IP$ and $\EQ$ queries---may 
be generated in a probabilistic manner.
Second, it can be assumed that all $\EQ$ queries occur only after 
all $\IP$ queries have been completed.
To see why this is so, start with an algorithm that interleaves 
$\IP$ and $\EQ$ queries, and modify it as follows.
Whenever an $\EQ$ query occurs before the end of the $\IP$ 
queries, the modified algorithm stores the value of the input 
to the query and proceeds as if the result were 0.
Then, at the end of the $\IP$ queries, each such deferred $\EQ$ 
query is applied.
The modified algorithm will behave consistently whenever the 
actual output of a deferred $\EQ$ query is 0, and also it will 
perform (albeit later) any $\EQ$ query where the output is 1.
Henceforth, we consider only algorithms with the above simplifications.

Now we describe a probabilistic procedure for constructing 
the black boxes that perform $\IP$ and $\EQ$ queries.
First, $a \in \01^n$ is chosen randomly according to the 
uniform distribution.
Then a set $S \subseteq \01^n$ is chosen randomly, uniformly 
subject to the condition that 
$|S| = (\half + \e)2^n$ (assuming that $\e2^n$ is an integer).
Then 
\begin{equation}
\IP(x) = \left\{
\begin{array}{ll}
a \cdot x \ & \mbox{if $x \in S$} \\
\overline{a \cdot x} & \mbox{if $x \not\in S$}
\end{array}
\right.
\end{equation}
and 
\begin{equation}
\EQ(x) = \left\{
\begin{array}{cl}
1 & \mbox{if $x = a$} \\
0 & \mbox{if $x \neq a$}.
\end{array}
\right.
\end{equation}

Consider an algorithm that makes $m$ $\IP$ 
queries.
If $m \ge \d n / \e^2$ then the theorem is proven.
Otherwise, since $\e \ge \sqrt{n}2^{-n/3}$, we have 
\begin{eqnarray}\label{bound_m}
m & < & \frac{\d n}{\e^2} \le \d 2^{2n/3}.
\end{eqnarray}

We proceed by determining the amount of information about $a$ that 
is conveyed by the application of $m$ $\IP$ queries.
Let $A$ be the $\01^n$-valued random variable corresponding to the 
probabilistic choice of $a \in \01^n$, and let $Y_1, Y_2, \ldots, Y_m$ 
be the $\01$-valued random variables corresponding to the respective 
outputs of the $m$ $\IP$ queries.
Let $H$ be the Shannon entropy function (see, e.g., \cite{CT91}). 
Then, for each $i \in \{1,2,\ldots,m\}$, 
\begin{equation}\label{recurrence}
H(A|Y_1,Y_2,\ldots,Y_i) = H(A|Y_1,\ldots,Y_{i-1}) 
- H(Y_i|Y_1,\ldots,Y_{i-1}) + H(Y_i|A,Y_1,\ldots,Y_{i-1}).
\end{equation}
Combining the above equations yields
\begin{equation}\label{main_equation}
H(A|Y_1,Y_2,\ldots,Y_m) = H(A) + \sum_{i=1}^m 
\left(H(Y_i|A,Y_1,\ldots,Y_{i-1}) - H(Y_i|Y_1,\ldots,Y_{i-1})\right).
\end{equation}
We shall now bound each term on the right side of 
Eq.~\ref{main_equation}.
Since the {\it a priori\/} distribution of $A$ is uniform, $H(A) = n$.
Also, since the entropy of a single bit is at most 1, 
$H(Y_i|Y_1,\ldots,Y_{i-1}) \le 1$ for all $i \in \{1,2,\ldots,m\}$.
Next, we show that, for all $i \in \{1,2,\ldots,m\}$, 
\begin{eqnarray}\label{inequality}
H(Y_i|A,Y_1,\ldots,Y_{i-1}) & \ge & 1 - (16 / \ln 2) \e^2.
\end{eqnarray}
To establish Eq.~\ref{inequality}, it is useful to view the set $S$ as 
being generated during the execution of the $\IP$ queries as follows.
Initially $S$ is empty, and when the first $\IP$ query is performed on 
some input $x$, $x$ is placed in $S$ with probability $\half + \e$ and 
in $\overline{S}$ with probability $\half - \e$.
The inputs to subsequent $\IP$ queries are also placed in either $S$ or 
$\overline{S}$ with an appropriate probability, which depends on how the 
inputs to previous queries are balanced between $S$ and $\overline{S}$.
After the execution of the first $i-1$ queries, the input to the 
$i^{\mbox{\scriptsize th}}$ query is placed in $S$ with probability 
\begin{equation}
\frac{(\half + \e)2^n - j}{2^n-(i-1)}, 
\end{equation}
where $j \in \{0,1,\ldots,i-1\}$ is the number of previous inputs to 
queries that have been placed in $S$.
Using Eq.~\ref{bound_m}, the above probability can be shown to lie 
between $\half - 2\e$ and $\half + 2\e$.
It follows that 
\begin{eqnarray}
H(Y_i|A,Y_1,\ldots,Y_{i-1}) 
& \ge & H(\half+2\e,\half-2\e) \nonumber \\
& = & - (\half+2\e)\log(\half+2\e) - (\half-2\e)\log(\half-2\e) \nonumber \\
& \ge & 1 - (16 / \ln 2)\e^2,
\end{eqnarray}
establishing Eq.~\ref{inequality}.
Now, substituting the preceding inequalities into 
Eq.~\ref{main_equation}, we obtain
\begin{equation}\label{entropy}
H(A|Y_1,\ldots,Y_m) \ge n - (16 / \ln 2) m \e^2.
\end{equation}
Intuitively, the $\IP$ queries yield information about the 
value of $A$ in terms of their effect on the probability distribution 
of $A$ conditioned on the values of $Y_1,\ldots,Y_m$.
Eq.~\ref{entropy} lower bounds the decrease in entropy possible.

From the conditions of the theorem, it can be assumed that, after 
the $\IP$ queries, $2^{n/2}$ $\EQ$ are performed.
The algorithm succeeds with probability at least $\d$ only if 
there exist $2^{n/2}$ elements of $\01^n$ whose total probability 
(conditioned on $Y_1,\ldots,Y_m$) is at least $\d$.
The maximum entropy that a distribution with this property 
can have is for a bi-level distribution, where $2^{n/2}$ elements 
of $\01^n$ each have probability $\d/2^{n/2}$ and $2^n - 2^{n/2}$ 
elements each have probability $(1-\d)/(2^n - 2^{n/2})$.
Therefore, 
\begin{eqnarray}
H(A|Y_1,\ldots,Y_m) 
& \le & H\big(%
\underbrace{\textstyle{\frac{\d}{2^{n/2}},
\ldots,\frac{\d}{2^{n/2}}}}_{2^{n/2}},%
\underbrace{\textstyle{\frac{1-\d}{2^n - 2^{n/2}},%
\ldots,\frac{1-\d}{2^n - 2^{n/2}}}}_{2^n-2^{n/2}}%
\big) \nonumber \\
& = & H(\d,1-\d) + \d \log(2^{n/2}) + (1 - \d)\log(2^n - 2^{n/2}) 
\nonumber \\
& < & 1 + \d n / 2 + (1 - \d) n \nonumber \\
& = & n - \d n / 2 + 1.\label{bi-level}
\end{eqnarray}
Combining Eq.~\ref{entropy} with Eq.~\ref{bi-level}, yields 
$ m > (\ln 2)(\d n - 2)/(32 \e^2) \in \Omega(\d n / \e^2)$, 
as required.
\qed\ee

We now provide definitions of {\it IP\/} and {\it EQ\/} queries 
in the quantum case in terms of unitary operations.
We do this in a manner that is sufficiently general so that, 
whenever an {\em implementation\/} of a more general {\it IP\/} 
or {\it EQ\/} query is given as a general quantum circuit consisting 
of elementary quantum gates and measurements, a unitary query 
corresponding to our definition can be efficiently constructed 
from it.

\begin{definition}\rm\label{quantum_IP}
A {\em quantum inner product\/} query (with bias $\e$) is a 
unitary transformation $\UIP$ on $n+m$ qubits, or its 
inverse $\UIPd$, such that $\UIP$ satisfies the following two 
properties:
\begin{enumerate}
\item
If $x \in \01^n$ is chosen randomly according to the uniform 
distribution and the last qubit of $\UIP\ket{x}\ket{0^m}$ 
is measured, yielding the value $w \in \01$, then 
$\Pr[w = a \cdot x] \ge \half + \e$.
\item
For any $x \in \01^n$ and $y \in \01^m$, the state of the first 
$n$ qubits of $\UIP\ket{x}\ket{y}$ is $\ket{x}$.
\end{enumerate}
\end{definition}
The first property captures the fact that, taking a query to be a 
suitable application of $\UIP$ followed by a measurement of the 
last qubit, Eq.~\ref{advantage} is satisfied.
Any implementation of a quantum circuit that produces an output 
that is $a \cdot x$ with probability on average $\half + \e$ 
can be modified to consist of a unitary stage $\UIP$ followed by a 
measurement of one qubit.
The second property is for technical convenience, and any unitary 
operation without this property can be converted to one that has 
this property, by first producing a copy of the classical basis 
state $\ket{x}$.
Moreover, given a circuit implementing $\UIP$, it is easy to 
construct a circuit implementing $\UIPd$.

\begin{definition}\rm\label{quantum_EQ}
A {\em quantum equivalence query\/}
is the unitary operation $\UEQ$ such that, for all 
$x \in \01^n$ and $b \in \01$,  
\begin{equation}
\UEQ\ket{x}\ket{b} = \left\{
\begin{array}{ll}
\ket{x}\ket{\overline{b}} & \mbox{if $x = a$} \\
\ket{x}\ket{b} & \mbox{if $x \neq a$,}
\end{array}
\right.
\end{equation}
where $\overline{b} = \neg b$.
\end{definition}

For the quantum GL problem, $a \in \01^n$ and information about $a$ 
in available only from quantum $\IP$ and $\EQ$ queries and the goal 
is to determine $a$.
We can now state and prove the result about quantum algorithms for 
the GL problem (which is similar to a result in \cite{CDNT99} in 
a different context).

\begin{theorem}\label{quantum_ub}\sl
There exists a quantum algorithm solving the GL problem with 
constant probability using $O(1/\e)$ 
$\UIP$, $\UIPd$ and $\UEQ$ queries in total.
Also, the number of auxiliary qubit operations used by 
the procedure is $O(n/\e)$.
\end{theorem}

\loud{Proof:}
The proof is by a combination of two techniques: the algorithm 
in \cite{BV93} for the exact case (i.e., when $\e = \half$), 
which is shown to be adaptable to ``noisy'' data in \cite{CDNT99} 
(with a slightly different noise model than the one that arises 
here); and amplitude amplification \cite{BH97,Gro96,BHMT00}.

Since $\UIP$ applied to $\ket{x}\ket{y}$ has no net effect on its 
first $n$ input qubits, for each $x \in \01^n$, 
\begin{equation}
\UIP\ket{x}\ket{0^m} = 
\ket{x}\left(\a_x \ket{v_x}\ket{a \cdot x} 
+ \b_x \ket{w_x}\ket{\overline{a \cdot x}}\right), 
\end{equation}
where $\a_x$ and $\b_x$ are nonnegative real numbers, and 
$\ket{v_x}$ and $\ket{w_x}$ are $m-1$ qubit quantum states.
If the last qubit of $\UIP\ket{x}\ket{0^m}$ is measured then the 
result is: $a \cdot x$ with probability $\a_x^2$, and 
$\overline{a \cdot x}$ with probability $\b_x^2$.
Therefore, since, for a random uniformly distributed $x \in \01^n$,
measuring the last qubit of $\UIP\ket{x}\ket{0^m}$ yields $a \cdot x$ 
with probability at least $\half + \e$, it follows that 
\begin{eqnarray}
{\textstyle{\frac{1}{2^n}}}
\sum_{x \in \01^n} \a_x^2 & \ge & \half + \e \\
{\textstyle{\frac{1}{2^n}}}
\sum_{x \in \01^n} \b_x^2 & \le & \half - \e.
\end{eqnarray}

\newpage

\begin{figure}[h]
\centering
\setlength{\unitlength}{0.5mm}

\begin{picture}(50,120)(0,0)
\end{picture}
\begin{picture}(165,120)(17,0)

\put(5,88){$\left\{ \mbox{\rule{0mm}{9mm}} \right.$}
\put(5,43){$\left\{ \mbox{\rule{0mm}{9mm}} \right.$}

\put(-17,88){\makebox(10,4){$n$ qubits}}
\put(-17,43){\makebox(10,4){$m$ qubits}}

\put(15,15){\line(1,0){10}}
\put(35,15){\line(1,0){50}}
\put(95,15){\line(1,0){70}}

\put(15,30){\line(1,0){25}}
\put(80,30){\line(1,0){20}}
\put(140,30){\line(1,0){25}}

\put(15,45){\line(1,0){25}}
\put(80,45){\line(1,0){20}}
\put(140,45){\line(1,0){25}}

\put(15,60){\line(1,0){25}}
\put(80,60){\line(1,0){20}}
\put(140,60){\line(1,0){25}}

\put(15,75){\line(1,0){10}}
\put(35,75){\line(1,0){5}}
\put(80,75){\line(1,0){20}}
\put(140,75){\line(1,0){5}}
\put(155,75){\line(1,0){10}}

\put(15,90){\line(1,0){10}}
\put(35,90){\line(1,0){5}}
\put(80,90){\line(1,0){20}}
\put(140,90){\line(1,0){5}}
\put(155,90){\line(1,0){10}}

\put(15,105){\line(1,0){10}}
\put(35,105){\line(1,0){5}}
\put(80,105){\line(1,0){20}}
\put(140,105){\line(1,0){5}}
\put(155,105){\line(1,0){10}}

\put(25,10){\framebox(10,10){$X$}}
\put(25,70){\framebox(10,10){$H$}}
\put(25,85){\framebox(10,10){$H$}}
\put(25,100){\framebox(10,10){$H$}}
\put(145,70){\framebox(10,10){$H$}}
\put(145,85){\framebox(10,10){$H$}}
\put(145,100){\framebox(10,10){$H$}}

\put(40,25){\framebox(40,85){$\UIP$}}
\put(100,25){\framebox(40,85){$\UIPd$}}

\put(85,10){\framebox(10,10){$Z$}}
\put(90,20){\line(0,1){10}}

\put(90,30){\circle*{3}}

\end{picture}
\caption{\small Quantum circuit $C$.}
\label{fig1}
\end{figure}
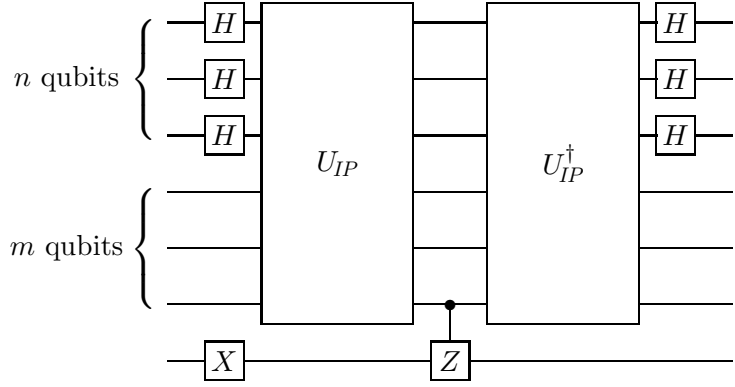

Now, consider the quantum circuit $C$ in Figure~\ref{fig1}.

\noindent We will begin by showing that $\bra{a,0^m,1}C\ket{0^n,0^m,0}$ is 
real-valued and 
\begin{equation}\label{overlap}
\bra{a,0^m,1}C\ket{0^n,0^m,0} \ge 2 \e, 
\end{equation}
which intuitively can be viewed as an indication of the progress that 
$C$ makes towards finding the string $a$.
To establish Eq.~\ref{overlap}, note that the operation $C$ can be 
decomposed into the following five operations:
\begin{enumerate}
\item
Operation $C_1$: Apply $H$ to each of the first $n$ qubits, and 
a NOT operation to the last qubit.
\item
Operation $C_2$: Apply $\UIP$ to the first $n+m$ qubits.
\item
Operation $C_3$: Apply a controlled-$Z$ to the last two qubits.
\item
Operation $C_4$: Apply $\UIPd$ to the first $n+m$ qubits.
\item
Operation $C_5$: Apply $H$ to each of the first $n$ qubits.
\end{enumerate}
Since $\bra{a,0^m,1}C\ket{0^n,0^m,0} 
= \bra{a,0^m,1}C_5 C_4 C_3 C_2 C_1\ket{0^n,0^m,0}$, 
the quantity $\bra{a,0^m,1}C\ket{0^n,0^m,0}$ is 
the inner product between state $C_3 C_2 C_1\ket{0^n}\ket{0^m}\ket{0}$ 
and state $C_4^{\dag} C_5^{\dag} \ket{a}\ket{0^m}\ket{1}$.
These states are 
\begin{eqnarray}
C_3 C_2 C_1 \ket{0^n}\ket{0^m}\ket{0} 
& = & C_3 C_2 {\textstyle{ \frac{1}{\sqrt{2^n}} }} \sum_{x \in \01^n} 
\ket{x}\ket{0^m}\ket{1} \nonumber \\ 
& = & C_3 {\textstyle{ \frac{1}{\sqrt{2^n}} }} \sum_{x \in \01^n} 
\ket{x}\left(\a_x\ket{v_x}\ket{a \cdot x} 
+ \b_x\ket{w_x}\ket{\overline{a \cdot x}}\right)\ket{1} \nonumber \\
& = & {\textstyle{ \frac{1}{\sqrt{2^n}} }} \sum_{x \in \01^n} 
\ket{x}\left(\a_x(-1)^{a \cdot x}\ket{v_x}\ket{a \cdot x} 
+ \b_x(-1)^{\overline{a \cdot x}}\ket{w_x}\ket{\overline{a \cdot x}}
\right)\ket{1} \nonumber \\
& = & {\textstyle{ \frac{1}{\sqrt{2^n}} }} \sum_{x \in \01^n} 
(-1)^{a \cdot x}\ket{x}\left(\a_x\ket{v_x}\ket{a \cdot x} 
- \b_x\ket{w_x}\ket{\overline{a \cdot x}}
\right)\ket{1}\label{A321}
\end{eqnarray}
and
\begin{eqnarray}
C_4^{\dag} C_5^{\dag} \ket{a}\ket{0^m}\ket{1} 
& = & C_4^{\dag} {\textstyle{ \frac{1}{\sqrt{2^n}} }} \sum_{x \in \01^n}
(-1)^{a \cdot x}\ket{x}\ket{0^m}\ket{1} \nonumber \\
& = & {\textstyle{ \frac{1}{\sqrt{2^n}} }} \sum_{x \in \01^n} 
(-1)^{a \cdot x}\ket{x}\left(\a_x\ket{v_x}\ket{a \cdot x} 
+ \b_x\ket{w_x}\ket{\overline{a \cdot x}}\right)\ket{1}.\label{A45}
\end{eqnarray}
It follows from Eq.~\ref{A321} and Eq.~\ref{A45} (and using the fact that 
$\bracket{x}{y} = 0$ whenever $x \neq y$) that 
\begin{eqnarray}
\bra{a,0^m,1}C\ket{0^n,0^m,0} 
& = & {\textstyle{ \frac{1}{2^n} }} \sum_{x \in \01^n} 
\left(\a_x^2 - \b_x^2\right) \nonumber \\
& \ge & (\half + \e) - (\half - \e) \nonumber \\
& = & 2 \e, 
\end{eqnarray}
which establishes Eq.~\ref{overlap}.

Note that Eq.~\ref{overlap} implies that, if $C$ is executed 
on input $\ket{0^n}\ket{0^m}\ket{0}$ ($= \ket{0^n,0^m,0}$) 
and the result is measured in the classical basis, then the 
first $n$ bits of the result will be $a$ with probability 
at least $|\bra{a,0^m,1}C\ket{0^n,0^m,0}|^2 \ge 4 \e^2$.
Therefore, if this process is repeated $O(1/\e^2)$ times, 
checking each result with an $\EQ$ query, then $a$ will be 
found with constant probability.
A more efficient way of finding the value of $a$ is to use 
{\em amplitude amplification\/} \cite{BH97,Gro96,BHMT00} using 
the transformation $C$ and its inverse $C^{\dag}$ in combination 
with $\EQ$ queries.
The procedure is to compute (for various values of $k$) 
\begin{equation}
(-C U_0 C^{\dag} \UEQ)^k C \ket{0^n,0^m,0} 
\end{equation}
(where $U_0 = I - 2\ket{0^n,0^m,0}\bra{0^n,0^m,0}$), measure the state, 
and perform an $\EQ$ query on the result.
Such a computation consists of $O(k)$ $\UIP$, $\UIPd$, and $\UEQ$ 
queries.
As shown in \cite{BHMT00}, if this is carried out for a suitably 
generated sequence of values of $k$, the expected total number of 
executions of $C$, $C^{\dag}$, and $\UEQ$ until a successful $\EQ$ 
query occurs is $O(1/\e)$.
This implies that $O(1/\e)$ $\UIP$, $\UIPd$, and $\UEQ$ are 
sufficient to succeed with constant probability.
\qed\ee

\section{Hard-predicates from one-way permutations}
\label{hard_predicates}

In this section, we give definitions pertaining to one-way permutations 
and hard-predicates (classical and quantum versions) and investigate 
the complexity of the reduction of Goldreich and Levin \cite{GL89} 
from the former to the latter.%
\footnote{The reduction makes sense for functions that are not 
permutations, but we restrict attention to permutations for 
simplicity.}

In the definitions below, when we refer to the {\em size\/} of a 
classical [quantum] circuit, it is understood to be relative to 
a suitable set of gates on one and two bits [qubits].
Quantum circuits compute unitary transformations on quantum 
states; however, they can also be adapted to take classical 
data as input and produce classical data as output.
For a quantum circuit $C$ acting on $m$ qubits, and 
$x \in \01^n$ (for $n \le m$), let $C_k(x)$ ($k \in \{1,\ldots,m\}$), 
denote the result of measuring the first $k$ qubits of 
$C\ket{x}\ket{0^{m-n}}$ in the classical basis.
The subscript $k$ may be omitted when the value of $k$ is clear 
from the context.

Intuitively, a quantum one-way permutation $f$ on $n$ bits is easy 
to compute in the forward direction but is hard to invert%
\footnote{The reversibility of quantum computations does not exclude 
this possibility \cite{CL97}.}.
For the former property, the standard requirement is that $f$ be 
computable by a uniform circuit of size $n^{O(1)}$ (though it is also 
possible to impose other upper bounds on the uniform circuit size).
To quantify the latter property, it is helpful to first make the 
following definition.
\begin{definition}\rm
A permutation $f : \01^n \rightarrow \01^n$ {\em is classically 
[quantumly] $(\d,T)$-hard to invert\/} if there is no classical 
[quantum] circuit $C$ of size $T$ such that $\Pr_a[C(f(a)) = a] \ge \d$.
\end{definition}
Now the standard requirement for the hard-to-invert condition is 
that $f$ is $(\d,T)$-hard to invert for all $\d \in 1/n^{O(1)}$ 
and $T \in n^{O(1)}$ (again, other bounds can be imposed).
It should be noted that, although it may be hard to determine 
$a$ from $f(a)$, it may not be hard to extract {\em partial 
information\/} about $a$ from $f(a)$.
For example, it is conceivable for a one-way permutation $f$ to 
have the property that {\em half\/} of the bits of $a$ can be 
efficiently determined exactly from $f(a)$.
It is also conceivable that each individual bit of $a$ is efficiently 
predictable from $f(a)$ with probability $\frac{3}{4}$.
The idea behind a hard-predicate \cite{BM84} is to concentrate the 
information that a one-way function ``hides'' about its input into a 
single bit.
Intuitively, $h : \01^n \rightarrow \01$ is a hard-predicate of 
$f$ if, given $a \in \01^n$, it is easy to compute $h(a)$; whereas, 
given $f(a)$ for randomly chosen $a \in \01^n$, it is hard to predict 
the value of the bit $h(a)$ with probability significantly better 
than $\half$.
One natural way of quantifying how well a circuit predicts the value 
of $h$ from the value of $f$ is by the amount that $\Pr_a[C(f(a)) = h(a)]$ 
exceeds $\half$.

The hard-predicate defined in \cite{GL89} is 
\begin{equation}
h(y,x) = y \cdot x, 
\end{equation}
(the inner product modulo two of $x$ and $y$), for 
$(y,x) \in \01^n \times \01^n$.
This is not a hard-predicate of $f$, but for a slightly 
modified version of $f$, as given in the following definition.
\begin{definition}\rm
For a permutation $f : \01^n \rightarrow \01^n$, let $\tilde{f}$ 
denote the permutation 
$\tilde{f} : \01^n \times \01^n \rightarrow \01^n \times \01^n$
defined as
\begin{equation}
\tilde{f}(y,x) = (f(y),x),
\end{equation}
for all $(y,x) \in \01^n \times \01^n$.
\end{definition}
Note that the cost of computing [inverting] $\tilde{f}$ is essentially 
the same as the cost of computing [inverting]~$f$.
Goldreich and Levin showed that if $f$ is one-way then $h$ is hard to 
predict from $\tilde{f}$.
Instead of quantifying how well a circuit predicts $h$ from $\tilde{f}$ as 
the amount by which $\Pr_{y,x}[C(\tilde{f}(y,x)) = h(y,x)]$ exceeds $\half$, 
we adopt a slightly more complicated definition.
This definition is related to the above, but is better suited for expressing 
the results in this section.
\begin{definition}\rm
A circuit $C$ {\em $(\d,\e)$-predicts $h$ from $\tilde{f}$} if 
\begin{equation}\label{predicts}
\Pr_y[\Pr_x[C(\tilde{f}(y,x)) = h(y,x)]\ge \half + \e] \ge \d.
\end{equation}
To explain Eq.~\ref{predicts} in words, call $y \in \01^n$ {\em $\e$-good\/} 
if $\Pr_x[C(\tilde{f}(y,x)) = h(y,x)]\ge \half + \e$ for that value of $y$.
Then then Eq.~\ref{predicts} is equivalent to saying that 
$\Pr_y[\mbox{$y$ is $\e$-good}] \ge \d$.
\end{definition}
The following lemma, which relates the two measures of prediction, 
is straightforward to prove by an averaging argument.
\begin{lemma}\sl
If $\Pr_{y,x}[G(\tilde{f}(y,x)) = h(y,x)] \ge \half + \e$ then 
$G$ $(\e/(1-\e),\,\e/2)$-predicts $h$ from $\tilde{f}$.
\end{lemma}
Note that, if 
$\Pr_{y,x}[G(\tilde{f}(y,x)) = h(y,x)] \ge \half + 1 / n^{O(1)}$ 
then $G$ $(1 / n^{O(1)},\,1 / n^{O(1)})$-predicts $h$ from~$\tilde{f}$.

The classical Goldreich-Levin Theorem can be stated as follows.
\begin{theorem}[\cite{GL89,Gol99}]\sl\label{classical_gl}
\ If $f : \01^n \rightarrow \01^n$ is classically $(\d/2,T)$-hard to 
invert then any classical circuit that $(\d,\e)$-predicts $h$ from 
$\tilde{f}$ must have size $\Omega(T \e^2 / n)$.
\end{theorem}
The proof of this theorem is essentially a reduction from the problem of 
inverting $f$ to the problem of $(\d,\e)$-predicting $h$.
One begins by assuming that a circuit $G$ of size $o(T \e^2 / n)$ 
$(\d,\e)$-predicts $h$ from $\tilde{f}$ and then shows that, 
by making $O(n/\e^2)$ calls to both $G$ and $f$ (plus some additional 
computations), $f$ can be inverted with probability $\d/2$ \cite{Gol99}.
The total running time of the inversion procedure is 
$o((n/\e^2)(T \e^2 / n)) = o(T)$, contradicting the fact that $f$ is 
$(\d/2,T)$-hard to invert.

Our quantum version of the Goldreich-Levin Theorem is the following.

\begin{theorem}\sl\label{quantum_gl}
If $f : \01^n \rightarrow \01^n$ is quantumly $(\d/2,T)$-hard to invert 
then any quantum circuit that $(\d,\e)$-predicts $h$ from $\tilde{f}$ must 
have size $\Omega(T \e)$.
\end{theorem}

\loud{Proof:}
As in the classical case, the proof is essentially a reduction from the 
problem of inverting $f$ to the problem of $(\d,\e)$-predicting $h$.
Let $b = f(a)$ be an input instance---the goal is to determine $a$ 
from $b$.
We will show how to simulate $\EQ$ and $\IP$ queries in this setting 
and then apply the bounds in Theorem~\ref{quantum_ub}.
It is easy to simulate an $\EQ$ query (relative to $a$) 
by making one call to $f$ and checking if the result is $b$.
Suppose that there exists a circuit $G$ of size $o(T \e)$ that 
$(\d,\e)$-predicts $h$ from $\tilde{f}$.
Thus, $\Pr_y[\Pr_x[G(\tilde{f}(y,x)) = h(y,x)] \ge \half + \e] \ge \d$.
Note that, with probability at least $\d$, $a$ is $\e$-good, in the sense 
that $\Pr_x[G(\tilde{f}(a,x)) = h(a,x)] \ge \half + \e$.
When $a$ is $\e$-good, computing $G(\tilde{f}(a,x)) = G(b,x)$ is simulating 
an $\IP$ query for $x$ (relative to $a$).
It follows from Theorem~\ref{quantum_ub} that $a$ can be computed with 
circuit-size $o((1/\e)(T \e)) = o(T)$ with success probability at least 
$\d/2$ (where $1/2$ is the success probability of the algorithm that 
finds $a$ when $a$ is $\e$-good and $\d$ is the probability that $a$ is 
$\e$-good to begin with).
This contradicts the $(\d/2,T)$-hardness of inverting $f$, thus $G$ cannot 
$(\d,\e)$-predict $h$ from $\tilde{f}$ and be of size $o(T \e)$.
\qed\ee

To conclude this section, we give a proof that Theorem~\ref{classical_gl} 
cannot be improved quantitatively assuming that it follows the structure 
of making calls to $f$ and to an algorithm $G$ that $(\d,\e)$-predicts $h$ 
from $\tilde{f}$.
More precisely, the setting is as follows.
For a permutation $f : \01^n \rightarrow \01^n$, information is available 
from two types of black-box queries: 
$f$-queries that evaluates $f$; and $G$-queries that $(\d,\e)$-predict 
$h$ from $\tilde{f}$.
More precisely, a $G$-query has the property that 
$\Pr_y[\Pr_x[G(\tilde{f}(y,x)) = h(y,x)] \ge \half + \e] \ge \d$.
A problem instance is $b \in \01^n$ (where $b = f(a)$ for a random 
$a \in \01^n$) and the availability of $f$-queries and $G$-queries.
The goal is to determine $a$ with probability $\d/2$ (say).
Let us refer to this as the {\em GL$^{\ast}$ problem\/} (related to but 
different from the GL problem defined in Section~\ref{query}).
From the proof of Theorem~\ref{classical_gl}, the classical GL$^{\ast}$ 
problem can be solved with $O(n/\e^2)$ $f$-queries and $G$-queries 
(and $O(n^2/\e^2)$ auxiliary operations \cite{Gol99}).
From the proof of Theorem~\ref{quantum_gl}, a quantum version of 
the GL$^{\ast}$ problem can be solved with only $O(1/\e)$ $f$-queries 
and $G$-queries (and $O(n/\e)$ auxiliary operations).
The next theorem essentially implies that the dilution factor $n/\e^2$ 
in Theorem~\ref{classical_gl} cannot be reduced for a reasonable range 
of values of $\e$.

\newpage

\begin{theorem}\sl
The classical GL$^{\ast}$ problem requires either $\Omega(2^{n/2})$ 
$f$-queries or $\Omega(n/\e^2)$ $G$-queries, whenever 
$\e \ge \sqrt{n}2^{-n/4}$.
\end{theorem}

\loud{Proof:}
The idea behind the proof is show that, starting with an algorithm that 
solves the GL$^{\ast}$ problem using $T_f$ $f$-queries and $T_G$ 
$G$-queries, it is possible to simulate each $f$-query with one $\EQ$ 
query and to simulate each $G$-query with one $\IP$ query and one $\EQ$ 
query.
The result is an algorithm that solves the GL problem defined in 
Section~\ref{query} with $T_G$ $\IP$ queries and $T_f + T_G$ $\EQ$ 
queries.
Then, applying the bound in Theorem~\ref{classical_lb}, yields the 
required lower bounds.

By a basic game-theoretic argument \cite{Yao83}, it suffices to consider 
deterministic algorithms, where the input data---embodied by $b$, $f$, and 
$G$---are generated in a probabilistic manner.
Let $a \in \01^n$ be chosen randomly, $f : \01^n \rightarrow \01^n$ be 
a random permutation (chosen uniformly among the $2^n!$ possibilities), 
and $b = f(a)$.
The function $G$ is generated with the following property: 
for any $y$, with probability at least $\d$, the condition 
$\Pr_x[G(\tilde{f}(y,x)) = h(y,x)] \ge \half + \e$ holds.
This property implies 
$\Pr_y[\Pr_x[G(\tilde{f}(y,x)) = h(y,x)] \ge \half + \e] \ge \d$.

The above probability distribution for $b$, $f$, and $G$ can be 
generated in a number of ways, including ways where the determination 
of parts of $f$ is deferred until the course of the execution of the 
algorithm solving the black-box problem.
To illustrate this, first consider an algorithm that uses only 
$f$-queries.
It is possible to generate $f : \01^n \rightarrow \01^n$ randomly, 
choose $a \in \01^n$ randomly, and set $b = f(a)$.
But this is stochastically equivalent to choosing $a \in \01^n$ 
randomly, $b \in \01^n$ randomly and then, whenever an $f$-query with input 
$x \in \01^n$ occurs, doing the following.
If $x = a$ then return $b$; if $x$ has already occurred as the input 
to an $f$-query then return the same value that was returned previously; 
otherwise, return a random element of $\01^n$ that is different from 
$b$ and from any values that have been returned from previous $f$-queries.
The above supposes that the value of $a$ is available.
If $b$ is available but information about $a$ is available only via 
$\EQ$ queries then, in the above procedure, checking whether $x = a$ 
can be replaced by performing the query $\EQ(x)$.
It is helpful to think about implementing the above process by building 
up a table of values of $f$, initially empty.
When an $f$-query with input $x$ occurs, an $\EQ$ query is performed.
If $\EQ(x) = 1$ then $b$ is returned; otherwise, if the table has a value 
$z$ in position $x$ then $z$ is returned; otherwise, a random $w \in \01^n$ 
that is different from $b$ and not in the table is inserted into position 
$x$ in the table and $w$ is returned.
This is the manner in which an $f$-query can be simulated by an $\EQ$ query.

In a similar spirit, we can show that $G$-queries can be incorporated into 
this scenario and simulated by $\IP$ queries and $\EQ$ queries.
Prior to the execution of the algorithm, a flag bit $s$ is set to 1 with 
probability $\d$ and to 0 with probability $1 - \d$.
Let the input to a $G$-query be $(y,x)$.
If $y = b$ and $s = 1$ then an $\IP$ query is performed and the result 
is returned.
If $y = b$ and $s = 0$ then a random bit is returned.
If $y \neq b$ and $y$ occurs in the table at position $z$ then 
$h(z,x)$ is returned.
If $y \neq b$ and $y$ does not occur in the table then $y$ is 
placed in a random empty position $z$ in the table for which 
$\EQ(z) \neq 1$ and $h(z,x)$ is returned.
In this manner, a $G$-query can be simulated by at most one $\IP$ 
query and one $\EQ$ query.

What results from the above is a method of converting an algorithm 
that solves the GL$^{\ast}$ problem with $T_f$ $f$-queries and $T_G$ 
$G$-queries with success probability at least $\d/2$ into 
one that solves the GL problem with $T_G$ $\IP$ queries and 
$T_f + T_G$ $\EQ$ queries with success probability $\d/2$.
Conditioned on $s=1$, this algorithm for the GL problem must succeed 
with constant success probability unless $T_f \in \Omega(2^{n/2})$.
Therefore, by the lower bounds in Theorem~\ref{classical_lb}, 
we have that $T_f \in \Omega(2^{n/2})$ or $T_G \in \Omega(n/\e^2)$, 
as required.
\qed\ee

\section{Quantum bit commitment from quantum one-way permutations}%
\label{bit_commitment}

In this section, we show how to use the quantum Goldreich-Levin Theorem 
to construct a quantum bit commitment scheme from a quantum one-way 
permutation.
\begin{definition}\rm
A permutation $f : \01^n \rightarrow \01^n$ is a {\em 
quantum one-way permutation\/} if: 
\begin{itemize}
\item
There is a uniform quantum circuit of size $n^{O(1)}$ that computes 
$f(x)$ from $x$.
\item
$f$ is quantumly $(\d,T)$-hard to invert for any $\d \in 1/n^{O(1)}$ 
and $T \in n^{O(1)}$.
\end{itemize}
\end{definition}

\begin{theorem}\label{thm4}\sl
If there exists a quantum one-way permutation 
$f : \01^n \rightarrow \01^n$ then there exists a bit [or qubit] 
commitment scheme that is perfectly binding and computationally 
concealing, in the sense that the committed bit cannot be predicted 
with probability $\half + 1/n^{O(1)}$ by a circuit of size 
$n^{O(1)}$.
\end{theorem}

\loud{Proof:}
From Theorem~\ref{quantum_gl}, it is straightforward to construct a 
quantum bit commitment scheme from Alice to Bob based on a 
one-way permutation $f$ as follows (where $h(y,x) = y \cdot x$).

\begin{description}
\item[Bit-commit]
Let $z \in \01$ be the bit to commit to.
Alice chooses $a, x \in \01^n$ randomly, and sets 
$c = z \oplus h(a,x)$.
Alice computes $b = f(a)$ and sends $(b,x,c)$ to Bob.
\item[Bit-decommit]
Alice sends $a$ to Bob.
Bob checks if $f(a) = b$ and rejects if this is not the case.
Otherwise, Bob accepts and computes $c \oplus h(a,x)$ 
as the bit.
\end{description}

Since $f$ is a permutation there is at most one {\em classical\/} 
value of $a$ that is an acceptable decommitment of Alice's bit.
This implies that the scheme is perfectly binding to Alice.
Note that the model could be relaxed to permit Alice to send 
quantum data to Bob, by adjusting Bob's protocol to immediately 
perform a measurement (in the classical basis) on any data that 
he receives from Alice.
There would be no advantage to Alice---she could not somehow 
``commit to more than one value'' by sending commitments in 
superposition.
This is because the adjusted protocol is equivalent to one where 
Alice performs the measurement herself on any data before sending 
it to Bob.

Theorem~\ref{quantum_gl} implies that the scheme is also 
computationally concealing, since any $n^{O(1)}$-size circuit 
that enables Bob to guess $z$ from $(b,x,c)$ with probability 
$\half + 1/n^{O(1)}$ can be converted to a $n^{O(1)}$-size 
circuit that inverts $f$ with probability $1/n^{O(1)}$, violating 
the fact that $f$ is one-way.

Finally, we explain how a {\em qubit\/} commitment scheme 
can be constructed using some of the ideas in \cite{AMTW00}.
Recall the standard notation for the Pauli matrices: 
\begin{equation}
X = \sigma_x = \left(
\begin{array}{cc}
0 & 1 \\ 1 & 0
\end{array}
\right)
\hspace*{10mm} \mbox{and} 
\hspace*{10mm} 
Z = \sigma_z = \left(
\begin{array}{cr}
1 & 0 \\ 0 & -1
\end{array}
\right).
\end{equation}
\begin{description}
\item[Qubit-commit]
Let $\ket{\psi}$ be the qubit to commit to.
Alice chooses $a_1, a_2, x_1, x_2 \in \01^n$ randomly, and 
constructs the state 
$\ket{\psi^{\prime}} = 
X^{h(a_1,x_1)}Z^{h(a_2,x_2)}\ket{\psi}$ 
and also computes $b_1 = f(a_1)$ and $b_2 = f(a_2)$.
Alice sends $(\ket{\psi^{\prime}},b_1,b_2,x_1,x_2)$ to Bob.
\item[Qubit-decommit]
Alice sends $a_1, a_2$ to Bob.
Bob checks if $f(a_1) = b_1$ and $f(a_2) = b_2$, rejecting if this 
is not the case.
Otherwise, Bob accepts and computes 
$Z^{h(a_2,x_2)}X^{h(a_1,x_1)}\ket{\psi^{\prime}}$ as the qubit.
\end{description}
Clearly, the scheme is perfectly binding.
Intuitively, the scheme is computationally concealing, because 
$h(a_1,x_1)$ and $h(a_2,x_2)$ ``look random'' to Bob.
If Bob can use his information to efficiently significantly distinguish 
between the qubit that he receives from Alice in the commitment stage 
and a totally mixed state (density matrix $\half I$) then this procedure 
can be adapted to distinguish between the pair of bits 
$r_1 = h(a_1,x_1)$ and $r_2 = h(a_2,x_2)$ and a pair of truly 
random bits, which would lead to a procedure that violated the 
result proven in Theorem~\ref{quantum_gl}.
\qed\ee

\section*{Acknowledgments}

We would like to thank Paul Dumais, Peter H\o yer, Dominic Mayers, and 
Louis Salvail for helpful discussions.

\end{document}